\documentclass[article]{aa}  
\usepackage{graphicx}
\usepackage{hyperref}                                                     
\usepackage{color}
\usepackage{txfonts}
\usepackage{placeins}
\newcommand{\kms}{km\,s$^{-1}$}

\newcommand{\Msun}{\,\mathrm{M}_\odot}

\begin{document}

   \title{Edge-On Disk Study (EODS) III: Molecular Stratification in the Flying Saucer Disk}

   \author{A. Dutrey \inst {1}
        \and O. Denis-Alpizar \inst{2}
        \and S. Guilloteau \inst{1}
        \and C. Foucher \inst{1}   
        \and S. Gavino \inst{3}
        \and D. Semenov \inst{4,5}
        \and V. Pietu \inst{6} 
        \and E. Chapillon \inst{1,6}  
        \and L. Testi \inst{3}
        \and E. Dartois \inst{7}
        \and E. DiFolco \inst{1}
        \and K. Furuya \inst{8}
        \and U. Gorti \inst{9}
        \and N. Grosso \inst{10}
        \and Th. Henning \inst{5}    
        \and J.M. Hur\'e \inst{1} 
        \and \'A. K\'osp\'al  \inst{11,12,5}
        \and F. Le Petit \inst{13}
        \and L. Majumdar \inst{14,15}
        \and R. Meshaka \inst{16}
        \and H. Nomura \inst{17}
        \and N.T. Phuong \inst{18}
        \and M. Ruaud \inst{9}
        \and Y.W. Tang \inst{19}
        \and S. Wolf \inst{20}
          }
  \institute{Univ. Bordeaux, CNRS, Laboratoire d'Astrophysique de Bordeaux, UMR 5804, F-33600 Pessac, France %
   \\
              \email{anne.dutrey@u-bordeaux.fr} 
    \and Departamento de Física, Facultad de Ciencias, Universidad de Chile, Av. Las Palmeras 3425, Ñuñoa, Santiago, Chile.
    \and Dipartimento di Fisica e Astronomia, Universit\`a di Bologna, via Gobetti 93/2, I-40190 Bologna, Italy
    Bologna, Italy
    \and ZAH/ITA, Heidelberg University, Albert-Ueberle-Str. 2, 69120 Heidelberg, Germany
     \and Max-Planck-Institut f\"{u}r Astronomie (MPIA), K\"{o}nigstuhl 17, D-69117 Heidelberg, Germany
    \and IRAM, 300 Rue de la Piscine, F-38046 Saint Martin d'H\`{e}res, France
    \and Institut des Sciences Moléculaires d'Orsay, CNRS, Univ. Paris-Saclay, Orsay, France 
    \and RIKEN Cluster for Pioneering Research, 2-1 Hirosawa, Wako-shi, Saitama 351-0198, Japan
    \and Carl Sagan Center, SETI Institute, Mountain View, CA, USA
    \and Aix-Marseille Univ, CNRS, CNES, LAM, Marseille, France 
    \and Konkoly Observatory, Research Centre for Astronomy and Earth Sciences, Konkoly-Thege M. út 15-17, 1121 Budapest, Hungary
    \and Institute of Physics and Astronomy, ELTE E\"otv\"os Lor\'and University, P\'azm\'any P\'eter s\'et\'any 1/A, 1117 Budapest, Hungary
    \and LUX, Observatoire de Paris, PSL Research University,
    CNRS, Sorbonne Universités, 92190 Meudon, France
    \and National Institute of Science Education and Research, Jatni 752050, Odisha, India
    \and Homi Bhabha National Institute, Training School Complex, Anushaktinagar, Mumbai 400094, India
    \and Astronomy Unit, School of Physics and Astronomy, Queen Mary University of London, London E1 4NS, UK
    \and National Astronomical Observatory of Japan, Division of Science, 2-21-1 Osawa, Mitaka, Tokyo 181-8588, Kanto Japan
    \and Viet. Nat. Space Center, Viet. Academy of Science and Technology, 18, Hoang Quoc Viet, Nghia Do, Cau Giay, Ha Noi, Vietnam
    \and Academia Sinica Institute of Astronomy and Astrophysics, 
    No.1, Sec.4, Roosevelt Rd, Taipei 106319, Taiwan, R.O.C
   \and Institut für Theoretische Physik und Astrophysik, Christian-Albrechts-Universität zu Kiel, Leibnizstraße 15, 24118 Kiel, Germany  
   }
   \date{\today}

 
  \abstract
{Investigating the vertical distribution of molecular content in protoplanetary disks remains difficult in most disks mildly inclined along the line of sight.
In contrast, edge-on disks provide a direct (tomographic) view of the 2D molecular brightness.} 
{We study the radial and vertical molecular distribution as well as the gas temperature and density by observing the Keplerian edge-on disk surrounding the Flying Saucer, a Class II object located in Ophiuchus.}
{We use new and archival ALMA data to perform a tomography of $^{12}$CO, $^{13}$CO, C$^{18}$O, CN, HCN, CS, H$_2$CO, c-C$_3$H$_2$, N$_2$D$^+$, DCN and $^{13}$CS. We analyze molecular tomographies and model data using the radiative transfer code \textsc{DiskFit}.}
{We directly measure the altitude above the mid-plane for each observed species. For the first time, we unambiguously demonstrate the presence of a common molecular layer and measure its thickness: most molecules are located at the same altitude versus radius. Beyond CO, as predicted by chemical models, the CN emission traces the upper boundary of the molecular layer, whereas the deuterated species (DCN and N$_2$D$^+$) resides below one scale-height. Our best fits from \textsc{DiskFit} show that most observed transitions in the molecular layer are thermalized  because their excitation temperature is the same, around $\sim$ 17-20\,K.}
{These long-integration observations clearly reveal a molecular layer predominantly located around 1-2 scale height, at a temperature above the CO freeze-out temperature. The deuterated molecules are closer to the mid-plane and N$_2$D$^+$ may be a good proxy for the CO snowline. Some molecules, such as CN and H$_2$CO, are likely influenced by the disk environment, at least beyond the mm dust disk radius. The direct observation of the molecular stratification opens the door to detailed chemical modeling in this disk which appears representative of T\,Tauri disks.}


   \maketitle
\nolinenumbers

\section{Introduction}

Studying the gas and dust content of protoplanetary disks orbiting young low-mass stars is fundamental to unveil the early phases of planet formation \citep{Oberg+2023}. 
After low and moderate angular resolution studies revealing the inventory of the most abundant molecules (CO, HCO$^+$, CN, HCN, CS, C$_2$H, H$_2$CO...) in T\,Tauri disks \citep[e.g.][]{Guilloteau+2016b}, 
view of these gas-rich, planet-forming disks drastically improved thanks  to ALMA. MAPS \citep{oberg+2021} studied the gas content of five 
large, relatively warm, disks orbiting T\,Tauri and Herbig\,Ae stars, unveiling radial structures and allowing more detailed analysis of chemistry \citep[e.g.][]{Guzman+2021,Oberg+2023}, as well as direct probe of the gas temperature \citep{Law+2021, Zhang+etal_2021}.

The inclination of a disk with respect to the line of sight is an important parameter. Pole-on disks, such as that of TW Hya, allow detailed investigations of the gas and dust radial distribution \citep[e.g.][]{Teague+2018,Andrews+2012,Huang+2018, Macias+2021}, but do not constrain the altitude at which the molecules reside inside the disk.
For inclined disks, the situation is more favorable thanks to the Keplerian gradient \citep{Dartois+etal_2003,Pinte+2018} and their analysis provide very interesting results on the molecular chemistry \citep{Bergin+2016,Semenov+2018,Guzman+2021, Kashyap+2024}.  Recent papers using more sensitive ALMA observations and upgraded methods have improved our knowledge on the understanding of the stratification of the vertical molecular layer \citep[e.g.][]{Paneque-Carreno+2022,Law+2023,Paneque+2023,Hernandez-Vera+2024,Law+2024,paneque-carreno+2024, Keyte+2024,Urbina+2024,Temmink+2025}. 
Nevertheless, the accuracy obtained still strongly varies with radius because the intensity decreases with radius ($>50-100$ au) and the sensitivity becomes limited. While the altitude of the emitting layer can be estimated, this is not the case for the vertical width of the molecular layer, particularly for optically thin and moderate opacity lines due to combined effects of inclination, density, temperature and velocity gradients along the line of sight.

Edge-on disks, in contrast, are ideal targets to reconstruct the full 2-D disk map by using its Keplerian rotation \citep{Dutrey+etal_2017,Flores+etal_2021}.

We present here new ALMA high angular resolution observations of the disk of the Flying Saucer \citep{Grosso+2003,Dutrey+etal_2017,Ruiz-Rodriguez+2021}, a Class II disk around a $0.60 \Msun$ T\,Tauri star  \citep{Simon+2019} located in Ophiuchus. An analysis of the continuum and CO data obtained in this project was presented in \citet[][hereafter Paper I]{Guilloteau+2025}. This second paper is dedicated to a first presentation of the extensive molecular data set focused on the molecular stratification (based on 25 detected transitions) accompanied by a simple but robust analysis. 
Forthcoming papers will target molecules of specific interest.

\section{Observations}

\subsection{Source presentation}

The edge-on Flying Saucer disk, located in Ophiuchus at $\sim$120 pc, was first imaged in the near-infrared domain by \citet{Grosso+2003}. 
The disk is not perfectly edge-on, and the scattered light morphology indicates that the Southern part is closest to us \citep{Grosso+2003}. 

Using ALMA data from project 2013.1.00387.S at 0.5'', \citet{Guilloteau+2016} reported the dust disk is settled, close to edge-on (inclination $> 85^{\circ}$), and has an outer dust radius of $\sim$ 180 au. 
They also measured the dust disk temperature because the emission of the bright CO background clouds is absorbed by the dust disk ($\sim7$\,K). \citet{Dutrey+etal_2017} made the first CO and CS tomographies showing their vertical molecular extent in the Flying Saucer. 
\cite{Ruiz-Rodriguez+2021} extended this to the analysis of CN emission available in the same data set, revealing the relative stratification of CO, CN and CS, and estimating the excitation conditions in the outer region ($\sim 240$ au) from the hyperfine intensity ratios of the N=2-1 transition.

The low gas dust and gas mid-plane temperature was confirmed by \citet{Guilloteau+2025} (Paper I), using higher angular data continuum and CO observations. Paper I also revealed the gas radial and vertical temperature gradients, and how the surrounding CO clouds affect the CO emission.

\subsection{Observations}
We use here the same ALMA Project 2023.1.00907.S (PI:\,O.Denis-Alpizar) as in Paper I. It uses one spectral setup in Band 6 and another one in Band 7, covering a large number of observed spectral lines (more than 20) at a spatial resolution as high as 0.18$''$ and spectral resolutions ranging from 0.040 to 0.2\,\kms (see Table\ref{tab:obs}). 
Observations and data reduction methods were already presented in Paper I, and are
summarized in Appendix \ref{app:obs}. Imaging parameters are described in Appendix \ref{app:imaging}.

We also included ALMA archival data from Projects 2013.1.00387 and 2013.2.00163.S that offered lower resolution observations of complementary spectral lines \citep{Guilloteau+2016,Dutrey+etal_2017,Simon+2019}.

\section{Results from the tomographies and models}

\subsection{Tomographies}

To present a synthetic view, we used the tomographic reconstruction method of \citet{Dutrey+etal_2017} that recovers the 2-D molecular brightness temperature distribution using the velocity structure (Appendix \ref{app:description-tomo} presents the method).

\begin{figure*}
\includegraphics[width=18.0cm]{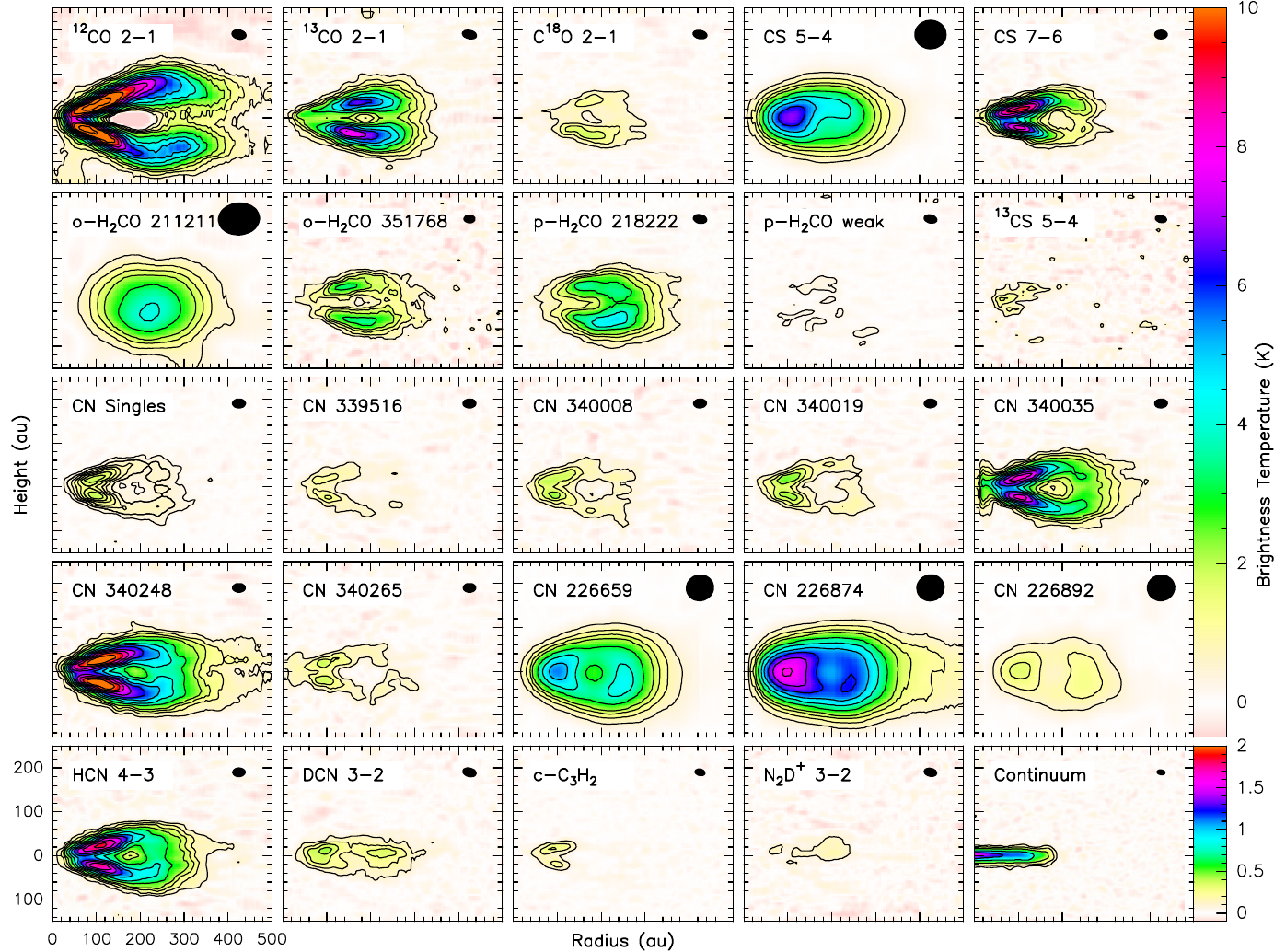} 
\caption{Tomographic view of the observed spectral lines, showing the mean brightness as a function of radius and height (both in au). Original beam sizes are indicated in upper right corner of each panel. CN hyperfine transitions are labeled by their frequency in MHz.
Contour levels are 0.5 to 2 by 0.5, 3 to 10 by 1, and 12 to 30 by 2 K (except for the weak para-H$_2$CO line, contour
at 0.25 K only). 
The panel labelled ``CN Singles'' shows a weighted average of 3 unblended lines are at 338516, 340008 and 340019 MHz, shown with a contour step
of 0.25\,K.
The continuum at 230 GHz is shown in the  bottom left panel, with contour levels  0.1 0.2 0.4 0.8  and 1.2 K.}
\label{fig:mean-tomo}
\end{figure*}

Figure \ref{fig:mean-tomo} displays the tomographies that clearly reveal the vertical stratification of the molecular emission in the disk, with little or no emission from the disk mid-plane in several cases. 

\begin{figure}
\includegraphics[width=9.0cm]{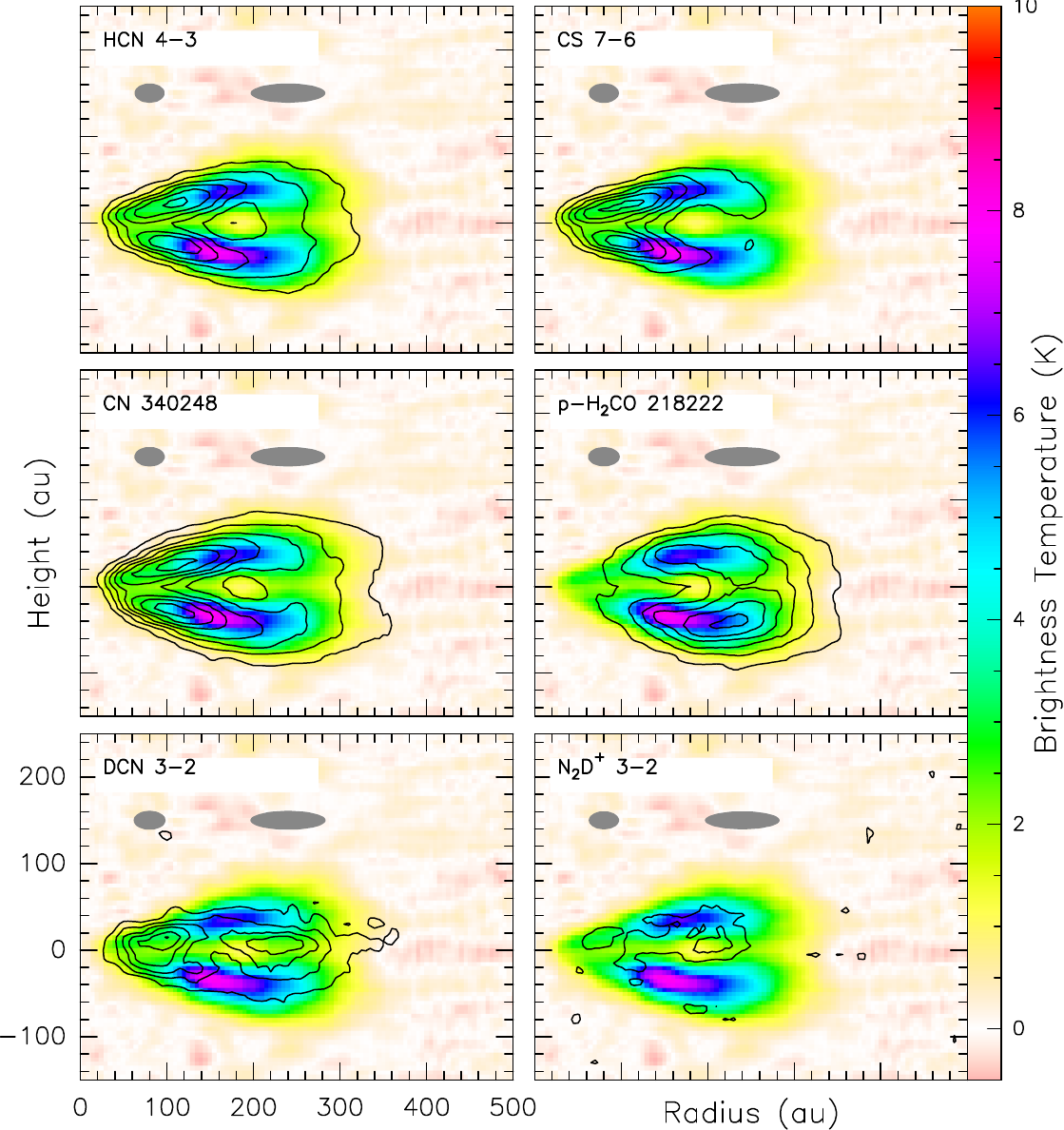}
\caption{$^{13}$CO tomography in false color with superimposition of the tomographies, in contours, of the HCN 4-3, CS 7-6, CN at 340.348 GHz, p-H$_2$CO at 218.222 GHz, DCN 3-2 and N$_2$D$^+$ 3-2. Contours are in steps
of 1.5\,K for HCN, CS, CN and p-H$_2$CO, 0.75\,K for DCN, and 0.375\,K for N$_2$D$^+$. The grey ellipses indicate the impact of line width on the effective beam at 80 and 240 au}
\label{fig:tomo-super}%
\end{figure}

Figure \ref{fig:tomo-super} shows the superimposition of the $^{13}$CO tomography (in false color) with those of important molecules such as CN, HCN, CS, H$_2$CO, DCN and N$_2$D$^+$ (in black contours). The figure clearly reveals differences in the vertical location of the various molecular emissions. Making cuts through the tomographies even better illustrates this.

Vertical cuts in tomographies at various radii (Fig.\ref{Fig:tomo-B7}) reveal that the vertical emission profiles can be well approximated by two Gaussian, although some top/bottom differences are observed. 
These profiles were hence fitted by two Gaussian, of same width and symmetrically placed below and above the disk plane for every radii.
The intensities of the two Gaussian were kept as free parameters and then averaged, when needed. 

\begin{figure*}
\includegraphics[width=18.0cm,height=7.25cm]{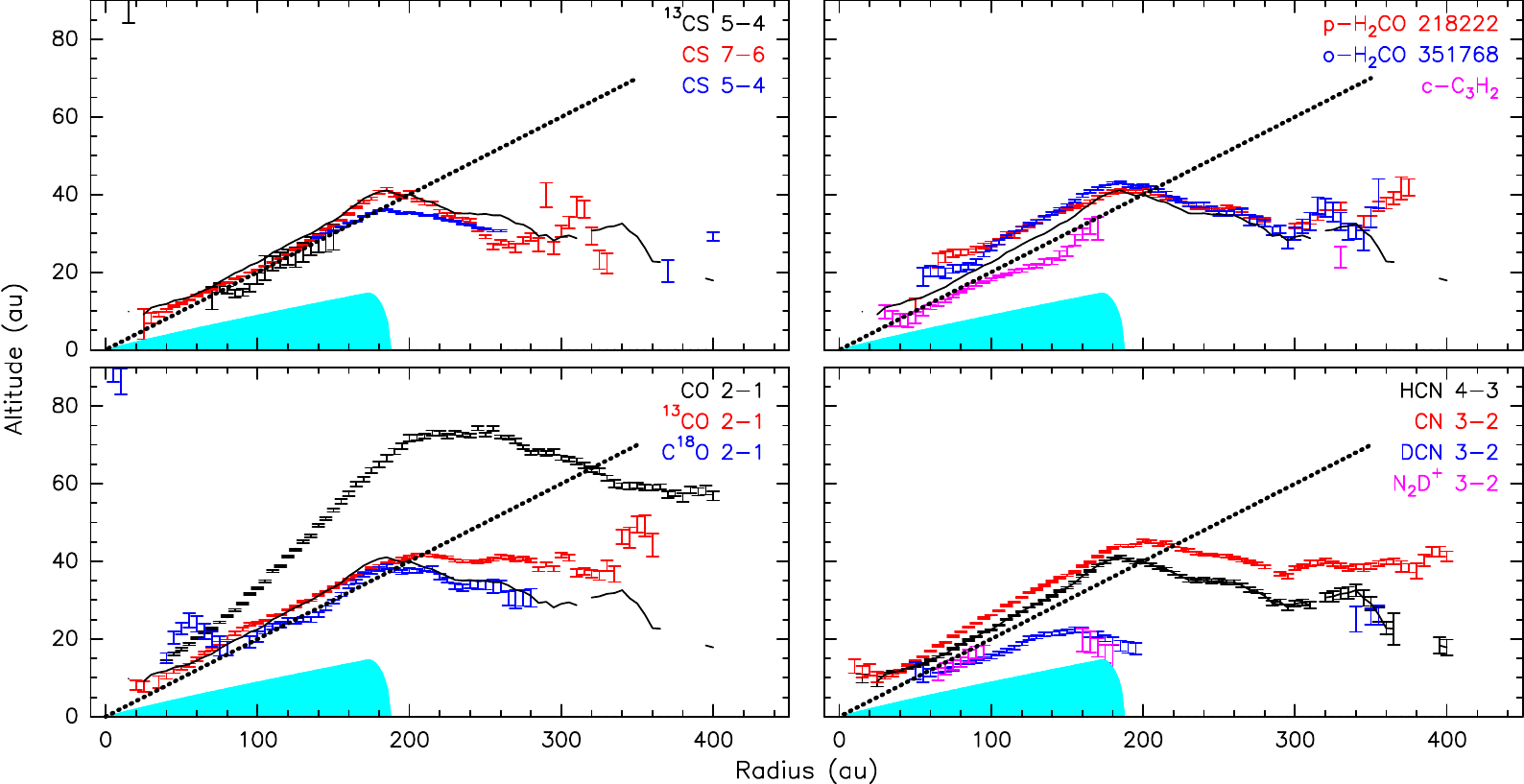}
\includegraphics[width=18.0cm,height=7.25cm]{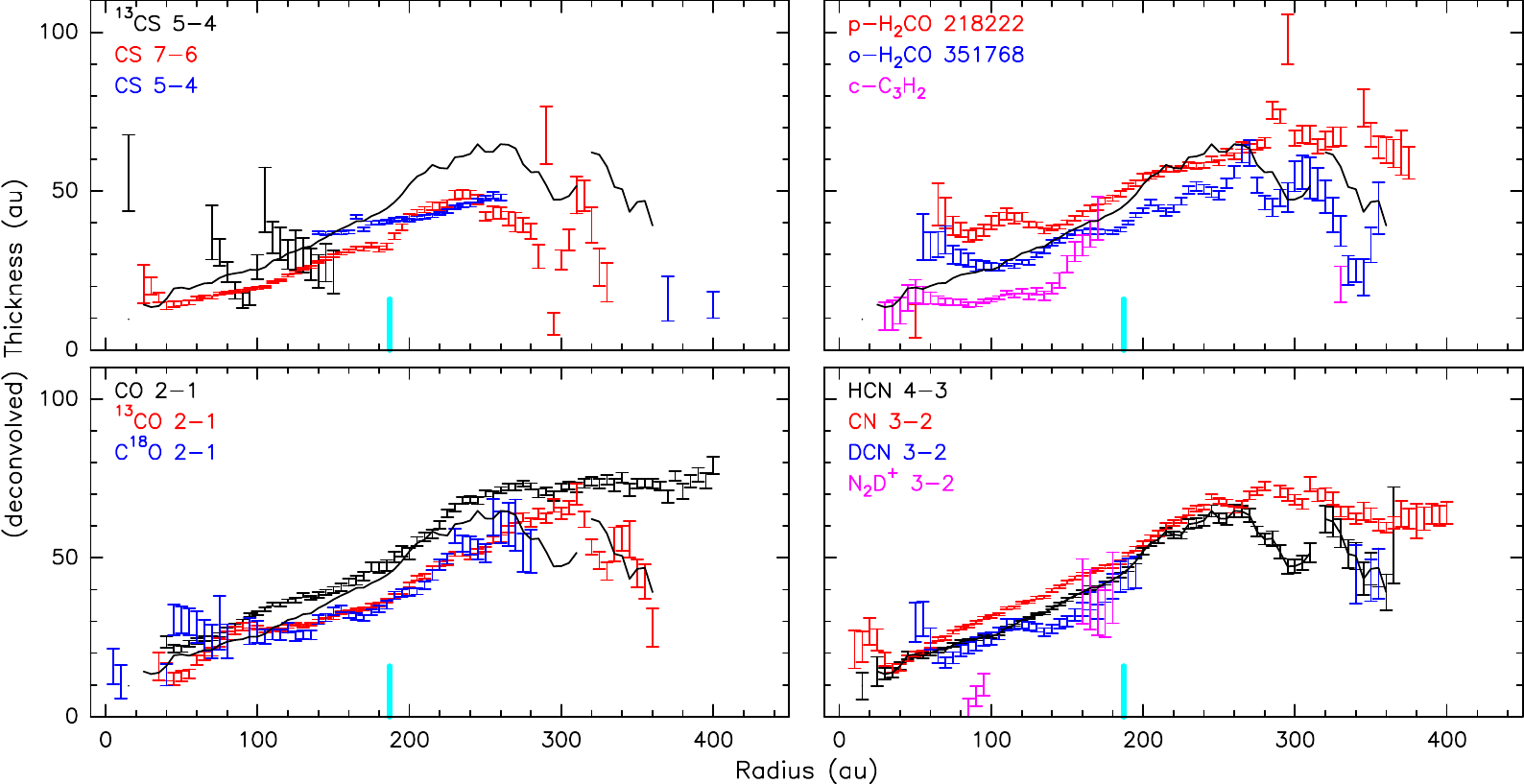}
\caption{Top: Altitude $A(r)$ of the molecular layer as a function of radius. The black curve is the HCN(4-3) altitude for comparison. The cyan region indicates the
the approximate size of the dust disk. The dotted line corresponds to z/r = 0.2.
Bottom: Deconvolved thickness of the molecular layer. The deconvolution is done assuming Gaussian shapes, using the Clean beam minor axis since synthesized beams are elongated almost parallel to the disk plane. The black curve is the HCN(4-3) thickness for comparison. The cyan bar marks the edge of the dust disk.}
\label{fig:height-tomo}%
\end{figure*}

The resulting fits with error bars are given in Fig.\ref{fig:height-tomo}, and Figs.\ref{fig:bright-tomo} to \ref{fig:width}. Fig.\ref{fig:height-tomo} shows the altitude and (deconvolved) thickness of the molecular emission.
Fig.\ref{fig:bright-tomo} shows the radial extent and intensity. 
Fig.\ref{fig:tomo-ratio} presents the south-to-north ratio that will be discussed in Sec.\ref{par:north-south}.
The measured widths (Fig.\ref{fig:width}) were deconvolved from the appropriate beam size for each line (bottom part of Fig.\ref{fig:height-tomo}). Comparisons of Fig.\ref{fig:width} with the bottom part of Fig.\ref{fig:height-tomo} show that the beam impact is small, except for the low resolution ($0.5''$) data where deconvolved widths are in agreement with those obtained at higher resolution. This clearly confirms that the thickness of the molecular layer is measured.  

\subsection{Disk model fitting} 
\label{sub:diskfit}
\begin{table*}
\caption{Model fitting results.}
\begin{tabular}{lllllllll}
\hline
 (1) & (2) & (3) & (4) & (5) & (6) & (7) & (8) & (9) \\
Parameter    
 & $R_\mathrm{int}$
 & $R_\mathrm{out}$
 & $H^m_0$
 & $Z_d$
 & $T_0$
 & $q$
 & $\Sigma_0$
 & $p$
 \\
Molecule 
 & (au)
 & (au)
 & (au)
 & 
 & (K)
 & 
 & cm$^{-2}$~~~(\%)
 & 
 \\
\hline
CO
&       24 $\pm$     1
&      296 $\pm$     4
&     19.1 $\pm$   0.1
&     0.71 $\pm$  0.36
&     26.6 $\pm$   0.1
&     0.28 $\pm$  0.01
&    1.7\,$10^{17}$ (4)
&     1.35 $\pm$  0.03
\\
$^{13}$CO
&       22 $\pm$     1
&      252 $\pm$     1
&     20.4 $\pm$   0.3
&     0.56 $\pm$  0.01
&     14.9 $\pm$   0.6
&     0.00 $\pm$  0.03
&   1.55\,$10^{15}$ (1)
&     0.07 $\pm$  0.07
\\
C$^{18}$O
&       25 $\pm$     9
&      246 $\pm$     2
& [    19.5 ]
&     0.46 $\pm$  0.13
& [    17.0 ]
& [    0.00 ]
&   2.16\,$10^{14}$ (1)
&     0.23 $\pm$  0.03
\\
CN(3-2)
&       44 $\pm$     4
&      292 $\pm$     1
&     19.3 $\pm$   1.3
&     0.52 $\pm$  0.01
&     18.6 $\pm$   0.2
&     0.46 $\pm$  0.01
&   1.12\,$10^{14}$ (1)
&     1.30 $\pm$  0.01
\\
CN(2-1)
&       55 $\pm$     1
&      292 $\pm$     1
&     18.1 $\pm$   0.1
&     0.59 $\pm$  0.01
&     19.6 $\pm$   4.8
&     0.31 $\pm$  0.01
&   1.27\,$10^{14}$ (1)
&     1.46 $\pm$  0.01
\\
HCN
&       34 $\pm$     1
&      257 $\pm$     2
&     21.1 $\pm$   0.1
&     0.46 $\pm$  0.01
&     17.3 $\pm$   0.5
&    -0.19 $\pm$  0.02
&   1.29\,$10^{13}$ (1)
&     2.73 $\pm$  0.03
\\
CS
&       34 $\pm$     2
&      253 $\pm$     1
&     18.1 $\pm$   0.1
&     0.63 $\pm$  0.01
&     20.7 $\pm$   0.1
&    -0.19 $\pm$  0.12
&    5.3\,$10^{13}$ (1)
&     3.21 $\pm$  0.02
\\
$^{13}$CS
&       46 $\pm$    16
&      205 $\pm$    17
& [    18.7 ]
& [    0.64 ]
& [    20.9 ]
& [   -0.07 ]
&   1.02\,$10^{12}$ (1)
& [    3.20 ]
\\
DCN
& [       5 ]
&      316 $\pm$    10
&     17.0 $\pm$   1.9
& [    0.00 ]
& [    10.0 ]
& [   -0.19 ]
&   1.58\,$10^{12}$ (1)
&     2.08 $\pm$  0.03
\\
N$_2$D$^+$
&       83 $\pm$     6
&      243 $\pm$    13
& [    19.0 ]
& [    0.00 ]
& [    10.0 ]
& [   -0.06 ]
&    8.9\,$10^{10}$ (33)
&     1.31 $\pm$  0.42
\\
o-H$_2$CO
&      106 $\pm$     2
&      282 $\pm$     2
& [    19.0 ]
& [    0.20 ]
&     14.6 $\pm$   0.2
& [    0.00 ]
&   1.44\,$10^{13}$ (1)
&     1.05 $\pm$  0.06
\\
p-H$_2$CO
&      134 $\pm$     1
&      305 $\pm$     1
& [    19.0 ]
& [    0.20 ]
&     12.4 $\pm$   0.3
& [    0.00 ]
&    7.5\,$10^{12}$ (2)
&     1.45 $\pm$  0.02
\\
c-C$_3$H$_2$
&       58 $\pm$     1
& [     300 ]
& [    19.0 ]
& [    0.00 ]
&     23.8 $\pm$   0.3
&     0.35 $\pm$  0.05
&   2.28\,$10^{13}$ (1)
&     2.77 $\pm$  0.06
\\
\hline
\end{tabular}
\tablefoot{Parameters within brackets were fixed. The physical parameters  $T_0$, $H^m_0$ and $\Sigma_0$ are referenced at $r_0 = 100$ au.  The scale height varies as $r^{1.25}$.  
For the surface density $\Sigma_0$ the error is quoted in percentage.
N$_2$D$^+$ and DCN column densities would be $\sim 30 \%$ smaller assuming a temperature of 17\,K. $Z_d$ mimicks the molecular depletion near the mid-plane, molecules being absent at altitude $z < z_d(r) = Z_d\,H(r)$, where $H(r)$ is the scale height of the molecule distribution (Col.4), see also Fig.\ref{fig:profiles}.} 
\label{tab:models}
\end{table*}

While the tomography unveils the geometry, it is not capable 
 to recover intrinsic physical parameters, such as molecular column densities. To do so we used the parametric radiative transfer code \textsc{DiskFit} \citep{Pietu+2007}.  Table \ref{tab:models} presents the best fit model for each molecule (apart from CN where the hyperfine structure allowed fitting the 3-2 and 2-1 transitions separately).

\begin{figure}
\includegraphics[width=9.0cm]{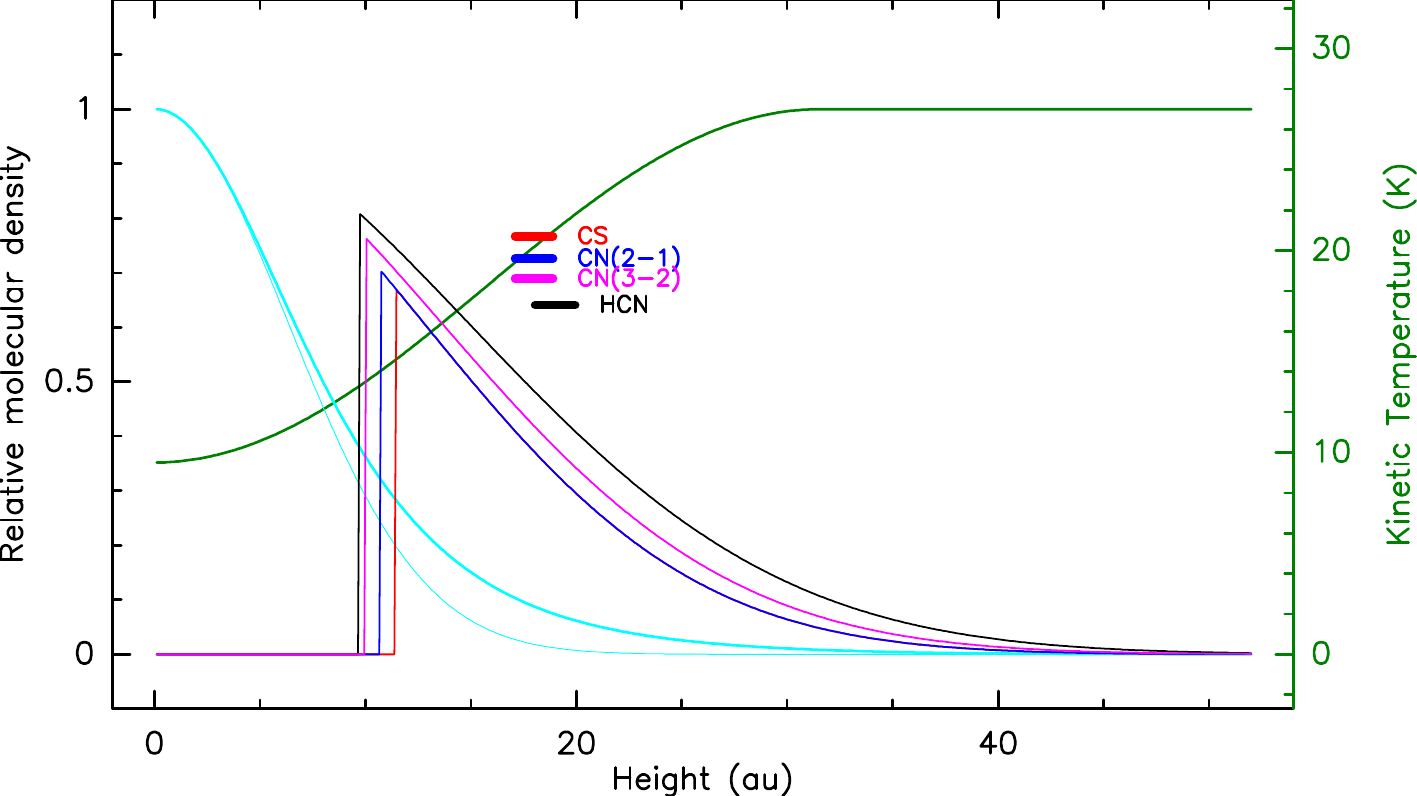}
\caption{Profiles resulting from \textsc{DiskFit} at 100\,au.
The green curve is the temperature profile, and the cyan curves indicate the H$_2$ density structure (thin: under isothermal assumption at the mid-plane temperature (10\,K), thick: using the full temperature profile) that were derived from the CO analysis by \citet{Guilloteau+2025}.
The other colored curves are the molecular density profiles for CS (red), HCN (black), CN 2-1 (blue) and 3-2 (magenta), using parameters from Table \ref{tab:models}. The corresponding horizontal bars indicate the measured temperatures and the density weighted
 average height of the emission.}
\label{fig:profiles}%
\end{figure}

The analysis with \textsc{DiskFit} is intended to provide the main properties of the 30-200 au region of the disk (where the dust disk extends), in particular the excitation temperature of the observed transitions and molecular surface densities. For this purpose, we adopted a simple Keplerian disk model, with power law radial distributions and sharp edges.
All radial quantities (temperatures, surface densities, scale heights) are power laws in the form $ x(r) = x_0 (r/r_0)^{-e_x}$, $r_o$ being the reference radius taken to be 100 au. 
The local linewidth $\delta V$ is assumed constant because of limited signal to noise. The model also incorporates continuum emission from dust, using the best fit model from Paper I. 

The molecular level populations are described by rotation temperatures that only depend on radius. The vertical molecular density profile is a Gaussian of independent (fitted) scale height
$$ n(r,z) \propto \exp(-(z/H^m(r))^2)$$ 
Each molecule $m$ has therefore a different apparent scale height $H^m(r)$ that characterizes the vertical location of the bulk emission of the given molecule. In addition, to represent the lack of emission near the mid-plane, molecules are absent at all heights where $z < Z_d\,H^m(r)$. For simplicity, here we assume that $Z_d$ does not depend on $r$. A graphic representation of the temperature and density distributions for key molecules is given in Fig.\ref{fig:profiles}. 
These approximations do not allow a proper representation of the outer regions ($r > 250$ au, where the emission reaches the mid-plane) but are appropriate for the bulk of emission, see Fig.\ref{fig:mean-tomo}. 
The apparent scale height and depletion level $Z_d$ are derived to reproduce the vertical brightness distribution, including the effect of line opacity. For $Z_d \sim 0.6$, most molecules reside around $H^m(r)$, which is thus comparable 
to the apparent height of molecular layer seen in Fig.\ref{fig:height-tomo}. An exception is CO, because of the large opacity of the 2-1 transition, but also of the increase in temperature in disk atmosphere. 
The molecule dependent scale heights $H^m$ should not be confused with the hydrostatic scale height of gas distribution, and Fig.\ref{fig:profiles} shows that the latter is about a factor two smaller than those of CN, HCN or CS.

For DCN and N$_2$D$^+$ which are located around the mid-plane in Fig.\ref{fig:mean-tomo} (and thus required  $Z_d \sim 0$), assuming a scale-height of 11 au (in agreement with the mid-plane temperature of 10 K) does not change the derived surface density within the current noise level. However, the vertical spread of DCN is best represented with an $H(r)$ similar to that of the other molecules.

For the temperature, we used the so-called LTE approximation, in which the molecular level population is controlled by a single temperature \citep[see][for the interpretation of this temperature]{Pietu+2007}. \textsc{DiskFit} then solves the radiative transfer equation along any line of sight by ray tracing, accounting for velocity gradients and taking the hyperfine structure of the spectral lines into account. Comparison to the data is made by comparing the interferometric visibilities resulting from the modeled intensity distribution to the observations.

The minimization is done through a Levenberg-Marquardt scheme, with multiple restarts (about 10) to avoid secondary minima. Errorbars are taken from the covariance matrix. The disk orientation, inclination and rotation velocity were derived from all strong lines, yielding consistent results (Table \ref{tab:geometry}). The kinematics was checked and found Keplerian for all lines. 

Line frequencies were taken from the CDMS data base. Details about individual molecules are given in Appendix \ref{app:fitting}.

Apart from CO, the molecules are vertically concentrated around a relatively narrow altitude range. Fig.\ref{fig:profiles} shows that the best model assuming a vertical temperature gradient (Paper I) samples temperatures from 15 to 20 K in this altitude range. Our independently fitted isothermal values falls exactly in this range, and are thus representative of the  average temperature of the molecular layer. Moreover, this isothermal assumption does not significantly affect the derived surface densities, because in the relevant temperature range (15-20 K) the brightness of the observed transitions in the optically thin regime only depends weakly on temperature \citep[see Fig.4 of ][]{Dartois+etal_2003}.

\begin{figure*}
\includegraphics[width=18.0cm]{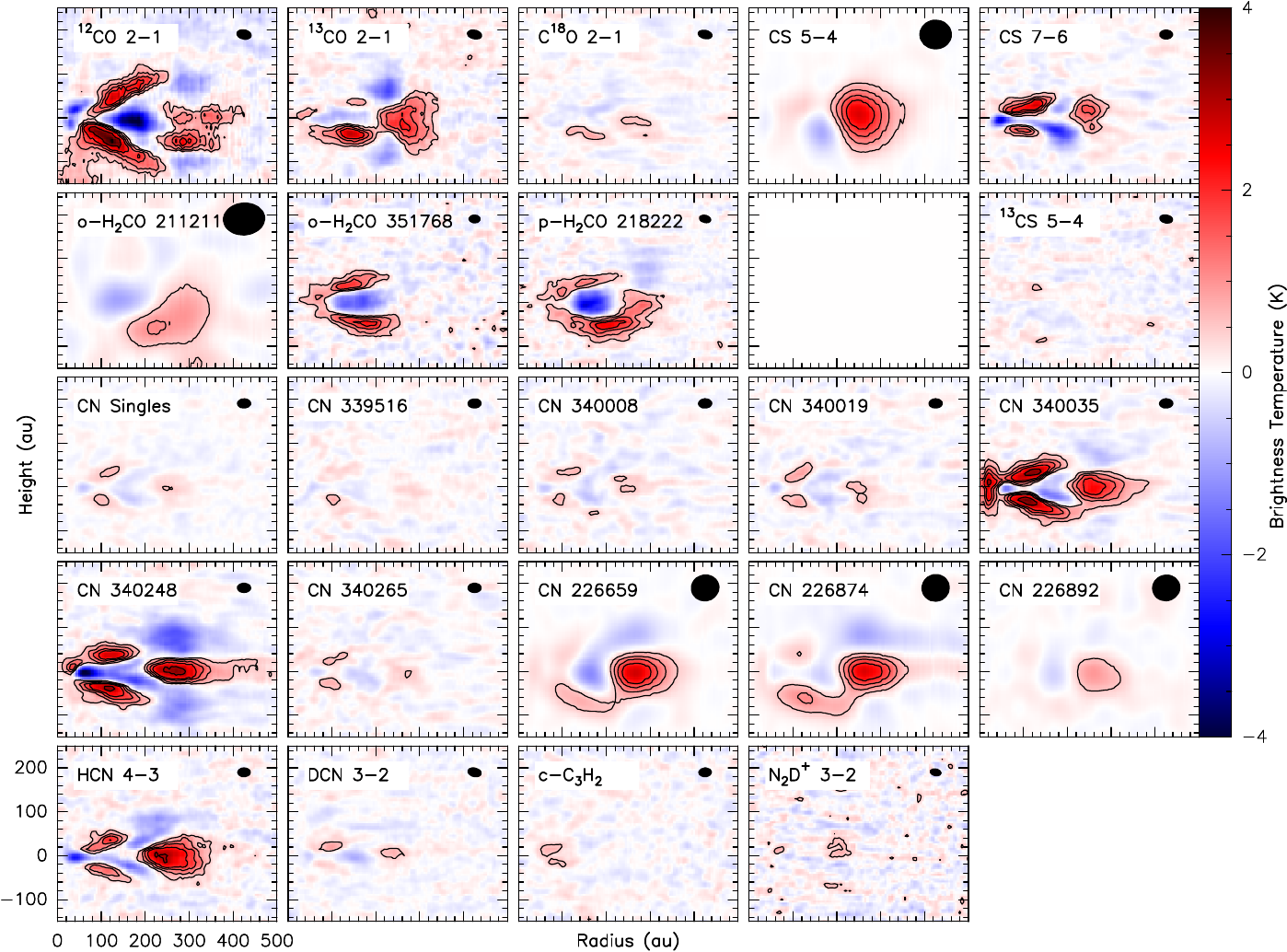}
\caption{Tomographies of the residuals of the \textsc{DiskFit} models. For each line, the best fit model visibilities  are subtracted from the observed ones, and imaged like for observations. The tomographic method is then applied on these residual data cubes. Contour spacings are 0.5\,K.}
\label{fig:tomo-residuals}%
\end{figure*}

Figure \ref{fig:tomo-residuals} shows the tomography of the residuals (the best fit model visibilities are subtracted from the observed ones, and imaged like for observations the tomographic method is applied on the data cubes of the residuals image) for lines with signal-to-noise ratio high enough. 

\section{Discussion}

\subsection{Vertical molecular stratification}

\paragraph{Molecular layer} 

Figures \ref{fig:mean-tomo} and \ref{fig:height-tomo} 
reveal that, with the exception of $^{12}$CO and the deuterated species, all molecules reside in a well-defined layer at an altitude scaling linearly with radius up to about 200 au.
Spatial resolution biases this altitude below $\sim 70$\,au because the two disk sides are no longer sufficiently separated (this bias is much larger for CS 5-4 or ortho-H$_2$CO at 211.211 GHz due to their lower $0.5-0.6''$ resolution). 
Figs.\ref{fig:tomo-super} and \ref{fig:height-tomo} clearly shows that the CN emission is more extended than those of HCN and CS, the latter being closer to the mid-plane. On the contrary, the DCN and N$_2$D$^+$ emissions clearly appear around the mid-plane.

The existence of such a well defined molecular layer has been predicted by many models over the last 20 years \citep[e.g.][]{Aikawa+2002}, see also \citet{,Dutrey+2014,Oberg+2023} for reviews.

The edge-on geometry provides for the first time the opportunity to 
directly measure the geometrical thickness of the molecular layer, as demonstrated in Fig.
\ref{fig:height-tomo} (see also Appendix \ref{app:description-tomo} and Fig.\ref{fig:width}). We find that the deconvolved layer thickness also scales approximately linearly with radius.

The \textsc{DiskFit} results (Table \ref{tab:models}) show that part of the disk mid-plane is devoid of molecules such as CN, CS, HCN, H$_2$CO, $^{13}$CO and C$^{18}$O. The apparent scale heights of the molecular distributions at 100 au ($H^m_0$, Col.4) are all similar (18-20 au) and there is no molecular emission below the altitude $z_d(r) \approx 0.5 H(r)$ for all molecules (including $^{13}$CO), except the deuterated species and c-C$_3$H$_2$. Finally, CS and $^{13}$CS are located at the same altitude. 

This is the first clear demonstration of a physical molecular layer common to many molecules, the width of the molecular layer being also properly measured thanks to the edge-on inclination, with  molecular vertical freeze-out likely happening over a few au. For comparison, \cite{Paneque+2023} reported in their sample of 7 inclined TTauri and Herbig Ae disks a molecular layer located at $z/r \leq 0.15$ (see their Figs.9 and 10). In our case, we clearly find $z/r \simeq 0.2$ inside a radius of 200 au and  $z/r \simeq 0.15 - 0.2$ beyond for all molecules with the exception of CO which is at higher $z/r \simeq 0.3 $ and the deuterated species which are at $z/r\simeq 0.1 - 0.12$.

\paragraph{Molecules near the mid-plane} 
The emission from N$_2$D$^+$, DCN and c-C$_3$H$_2$ is located closer to the mid-plane (Fig. \ref{fig:tomo-super} and \ref{fig:height-tomo}) with the best fits converging toward Z$_d = 0$. 
Following \citet{Willacy+2007}, DCN is formed around $\sim$ 10\,K via ion-molecules reactions. 
We therefore assumed a temperature of 10\,K to derive the column density, checking that this value (consistent with the mid-plane temperature found in Paper I) does not affect the other parameters, in particular $Z_d$. We only find a difference in surface density $\Sigma$ of 30\%.

\citet{Majumdar+2017} and \citet{Aikawa+2018} found that the formation of N$_2$D$^+$ is very efficient at low temperatures when CO is frozen onto grains, being then enhanced in the cold mid-plane, and thus may better trace the CO snow line than N$_2$H$^+$. 
\cite{cataldi+2021} reported that the inner radius of the N$_2$D$^+$ column density profile roughly corresponds to the location of the CO snowline in most sources of project MAPS (except GM\,Aur because of a low S/N ratio). 
In  the Flying Saucer, despite limited S/N ratio, the N$_2$D$^+$ emission coincides with the location of the hole observed in CO (Fig.\ref{fig:mean-tomo} and \ref{fig:tomo-super}) at radius $\sim$ 200 au. 

\paragraph{South-to-North asymmetry}
\label{par:north-south}
In a flared disk, not exactly edge-on, the nearest side will be seen at higher inclination angle that the most distant one. For partially optically thick lines, the outer (colder) regions will self-absorb the emission coming from innermost (warmer) zones. On the opposite, the furthest side is seen more face on, allowing to peer directly into the warmer regions. Hence, the furthest side should appear warmer. This was mentioned for the first time by \cite{Guilloteau-Dutrey1998} for data in DM Tau, and has since been used in many disks to recover the full disk orientation. The specific case of edge-on disks not exactly at 90 degrees is modeled in \cite{Dutrey+etal_2017}.

CO, CS and the inner region in HCN exhibit a South-to-North asymmetry (Fig.\ref{fig:tomo-ratio}) consistent with the known disk orientation, with a brighter Northern side. 
However, for molecules with low or moderate opacity the Southern side appears brighter, at least beyond the dust disk radius in the 150-200 to 300 au range. This is more pronounced at the lowest opacities (C$^{18}$O, H$_2$CO) than for CN (and HCN in the outer part). 
There is no obvious explanation for this intrinsic asymmetry. 
It may result from a difference  in the disk illumination, caused e.g. by an unseen disk warp, or by an asymmetry in the external UV field. 
In this respect, \citet{Dutrey+etal_2017} noted that the Flying Saucer is located on the eastern side of the $\rho$\,Oph cloud while most B stars are located on the western side, thus allowing dense clouds to absorb most of their UV radiation.

\subsection{Radial distribution}
\label{sub:radial}
The radial distribution can be roughly separated in four parts\,: an inner cavity, a first molecular ring, a large gap (in the mid-plane) and a second molecular ring (or second emission peak) located beyond the mm dust outer radius. The sizes of the cavities and gaps vary depending on the molecules while the second molecular ring is not always seen.

\paragraph{Inner radii}
\cite{Dutrey+etal_2017} reported an upper limit on a possible central hole of $\sim$\,15 au for CO and $\sim$\,25 au for CS based on the position-velocity diagrams.
This is roughly consistent with the values we determined for the CO and CS lines using \textsc{DiskFit}. 
The continuum does  not exhibit a central cavity and is not optically thick enough to hide molecules 
with an average dust opacity of $\sim$\,0.2 (Paper I).
Moreover, the size of the central cavity is slightly larger for CN ($\sim$\,50 au), suggesting that chemical effects may play a role in creating these inner depressions.

The H$_2$CO and N$_2$D$^+$ emissions exhibit very large inner cavities (Table \ref{tab:models} and Fig.\ref{fig:bright-tomo}).  
H$_2$CO emission extends from radii 130 au to 300 au, i.e. most of the  H$_2$CO emission comes from beyond the dust disk. N$_2$D$^+$ emission has an inner radius of 80 au and its maximum intensity at the dust disk outer radius ($\sim$\,200 au) as the very large gap seen in the CO emission. 
The very large gap is also clearly seen in Fig.\ref{fig:tomo-super}. 

\paragraph{Rings in the molecular layer} 
In the molecular layer, the radial distribution of CO emission cannot be simply interpreted due to confusion with the surrounding clouds, see Paper I for details, but the $^{13}$CO and C$^{18}$O emissions (essentially free of confusion) exactly peak in the same radius range, 150-180 au. 

In contrast, the maximum intensity of the optically thin $^{13}$CS 5-4 emission radially coincides with the radial location of the maximum intensity of the CS 7-6 emission ($\sim 120$ au, Fig.\ref{fig:bright-tomo}). CN 3-2 and HCN 4-3 emission also peaks at a radius of $\sim$ 120 au, like CS.  DCN emission is prominent at radii 70-100 au and then 200-220 au. 
This suggests the existence of at least two resolved rings (see also Fig.\ref{fig:bright-tomo}), as already observed in less inclined disks \citep[e.g.][]{Oberg+2023}. \cite{cataldi+2021} also reported the presence of DCN beyond the radius of the dust disk of mm in IM\,Lupi and GM\,Aur. The c-C$_3$H$_2$ emission peaks at 70-80 au, as the first peak of DCN emission. 
Unlike other species, the H$_2$CO emission exhibits two maxima located beyond a radius of 150 au. The outer radius of the continuum being at $\sim$ 180 au, most of the  H$_2$CO emission extends outside the dust disk.

In conclusion, with the exception of CO lines, there are at least two bright molecular areas, an inner region extending from about 50 to 150 au, an outer ring (secondary emission peak) peaking beyond the mm dust outer radius at 200-220 au (see also Fig.\ref{fig:tomo-super}).

\subsection{Temperature in the molecular layer}

Table \ref{tab:models} shows that the temperature in the molecular layer is about 17-20 K. $^{13}$CO is partially optically thick and provides a temperature measurement in the radius range 150--200 au. 
Our vertically isothermal analysis provides results in excellent agreement with those obtained in Paper I
with a temperature model that followed the vertical prescription by \citet{Dutrey+etal_2017}. 
This justifies our simplified assumption, with only $^{12}$CO probing the upper most layers.
This is also clearly demonstrated by Fig.\ref{fig:profiles} which compares the average temperature derived from this paper with the temperature derived from CO using the more physical description of Paper I.

The derived temperature is always slightly above the CO freeze-out temperature of 17 K (\citealt{Qi+2013}, although there are variations on the exact value for the CO freeze-out temperature, e.g. \citealt{Qi+2019}). Nevertheless, the derived temperature can be considered as representative of the gas temperature in the bulk of the molecular layer. 

Similarly, \citet{Foucher+2025} reported, for the molecular layer of the cold edge-on disk of SSTTau 042021, a temperature (derived from CO and $^{13}$CO ALMA data) that is also almost radially constant and also slightly above the CO freeze-out temperature of 17 K.

\paragraph{Radial dependence}
The temperature appears also almost constant in the radius range 50--200 au, the value of exponent $q$ being close to 0 for CS, HCN and CO isotopologues. Our dual line analysis (J=5-4 and J=7-6) of CS yields a temperature that is in excellent agreement with that found by \citet{Dutrey+etal_2017} based only on the transition of 5-4.

\paragraph{The case of CN}
For CN, we were able to measure the temperature independently for each transition thanks to the hyperfine structure that directly constrains the line opacities (see Appendix \ref{app:fitting} for details). The same value of 19--20 K is found for the N=2-1 and N=3-2 transitions. \cite{Ruiz-Rodriguez+2021} used the same CN 2-1 data at $0.5''$ resolution to derive the temperature in the outer disk, beyond 210 au, and found T\,$\approx 12$\,K in the midplane. Although our
model is based on measurements at smaller radii, our extrapolated value at 250 au is $\sim 13$\,K.

\subsection{Surface density in the molecular layer}

CO isotopologues show flat surface density profiles ($p \simeq 0$). 
The $^{13}$CO/C$^{18}$O ratio of $\sim$\,8, consistent with the local elemental isotopic ratios \citep{Wilson+1994}, suggests a relatively limited impact of selective photo-dissociation and fractionation in the molecular layer for CO \citep{Langer+1980, Miotello+2014}. 
The derived CS/$^{13}$CS ratio of about 50 also excludes strong fractionation in this molecule \citep{Wilson+1994,rodriguez-baras+2021}.

HCN, CS and c-C$_3$H$_2$ lines exhibit a steep decline with radius (exponent $p$). This may be due in part to excitation conditions, at least for CS and HCN. 
The surface densities of CN and H$_2$CO have an intermediate behavior ($p\sim 1.5 $) since their emission extends beyond the radius of the dust disk. This may be related to significant changes in chemistry and illumination. 
The CN surface density, $\sim 1.2 \cdot 10^{14}$cm$^{-2}$ at radius 100 au, and CN/HCN ratio ($\sim 10$) are also similar to the values quoted by \citet{Bergner+2021} in IM Lupi and GM Aur. 

Finally, a comparison of the results from Table \ref{tab:models} with the surface densities derived from 30 T\,Tauri disks by \citet{Guilloteau+2016} clearly shows that the Flying Saucer disk exhibits gas properties representative of those of T\,Tauri disks. 

\subsection{Departure from a simple model}
\label{sub:outer}
Fig.\ref{fig:tomo-residuals} shows the tomographies of the residual emission after subtraction of our best fit \textsc{DiskFit} model from the data. 
The CO emission cannot be simply discussed due to the confusion of the CO cloud.
$^{13}$CO is also affected to some extent, which may partially explain its anomalous South/North intensity ratio. 
However, other molecules are free of confusion. Beyond 200 au, CN, HCN, CS, and to a lesser extent $^{13}$CO  clearly exhibit emission close to the disk mid-plane from the outer ring.
As our disk model assumes the emission height varies as a power law, it is unable to properly represent the regions beyond about 200\,au, where the altitude of the molecular layer starts decreasing (see Fig.\ref{fig:height-tomo}). 

These residuals also reveal that CN appears vertically more extended than in our simple model, with significant residuals above the common molecular layer. Such residuals are much fainter in HCN, although fit results in Table \ref{tab:models} do not indicate significant differences in the altitude of CN and HCN. 
This indicates that CN, although predominantly located in the molecular layer, extends to higher altitudes, as expected because of its sensitivity to UV radiation. 
A more complex approach is needed to quantify this behaviour.
CS displays an intermediate behaviour between CN and HCN in this respect. 
The H$_2$CO location is also different than our simple assumption, being both vertically and radially more extended. 
These residuals indicate that our assumed vertical profiles, displayed in Fig.\ref{fig:profiles}, may require a more pronounced tail at high altitudes instead of our assumed Gaussian drop off.

\section{Conclusion and Perspectives}

For the first time, we clearly show that most molecules (apart from the deuterated species) reside in a common molecular layer and we are able to measure its geometrical thickness. 
Furthermore, the transition between the depleted mid-plane and the lower edge of this layer seems to occur over a few au (see Fig.\ref{fig:profiles}).
This unique data set provides new insights that deserve more sophisticated thermo-chemical modeling to be understood, {as illustrated by Figs.\ref{fig:mean-tomo} and \ref{fig:tomo-residuals}. We emphasize some important points here. 
\begin{itemize}
\item 
We find that the molecular layer is at z/r $\sim$ 0.2, larger than in the sample studied by \cite{Paneque+2023} (z/r $\sim$ 0.1), perhaps due to the lower disk temperature as the central star is of lower mass than in their sample. This needs to be confirmed by studying other edge-on disks. 
\item The molecular layer has a relatively shallow radial temperature profile, $\sim 17-20$ K at radius 100 au, only slightly above the CO freeze-out temperature. 
\item As predicted by astrochemical models \citep[e.g.][]{Aikawa+2018}, the deuterated species are clearly located closer to the mid-plane, but likely at a lower altitude than predicted for TW Hydra \citep{Romero-Mirza+2023}.  
The N$_2$D$^+$ emission has a low S/N ratio but its extent suggests that N$_2$D$^+$ may be a good tracer of the CO snowline. 
Comparing N$_2$H$^+$ and N$_2$D$^+$ distributions would be invaluable. 
\item As visible from Figs.\,\ref{fig:tomo-residuals} and \ref{Fig:tomo-B7}, although CN has a scale height similar to the other molecules, the CN layer appears slightly vertically more extended than most of the other molecules except CO \citep[see also][]{Paneque-Carreno+2022} probably because of its sensitivity to UV radiation \citep{Cazzoletti+2018,Nomura+2021, ruaud+2021}. 
\citet{Yoshida+2024} and \citet{Paneque+2023} observed the same behavior in the case of TW Hydra and Elias 2-27, respectively. For Elias 2-27, it appears compatible with predictions from thermo-chemical models. Large CN radial extents (in part beyond mm dust disks) are also seen by \citet{Bergner+2021} for the five sources of the MAPS project.
\item The H$_2$CO distribution has a large central cavity and
peaks beyond the dust disk edge, a situation similar 
to that found by \citet{Guzman+2021, Hernandez-Vera+2024} for AS\,209, HD\,163296 and IM\,Lupi. The ortho/para ratio is around 2, consistent
with the low temperature of 12-14\,K. The different
slopes ($p$) indicate that this ratio may increase with radius.
This may reflect the formation of H$_2$CO by two different routes, a cold one on grain surfaces \citep{Fayolle+2016}, and a gas-phase one as for DM\,Tau \citep{Loomis+2015}. 
\item The c-C$_3$H$_2$ emission steeply
decreases with radius, like $^{13}$CS, but appears somewhat below the molecular layer.
The latter point is also reported by \citet{Paneque+2023} in the case of the warmer disk of HD\,163296.
\item Finally, several molecular emissions have a secondary emission peak or outer ring beyond the millimeter dust outer radius, suggesting that photo-desorption may play an important role for many molecules to replenish the gas phase in cold disks around T\,Tauri stars.
\end{itemize}
These points clearly deserve deeper observational and model studies to confirm the observed behaviors on more sources in different molecules.

\begin{acknowledgements}
This work was supported by the Thematic Actions PNPS and PCMI of INSU Programme National “Astro”, with contributions from CNRS Physique \& CNRS Chimie, CEA, and CNES. 
%
NTP acknowledges support from the Vietnam Academy of Science and Technology under grand number VAST 08.02/25-26.
Y.W.T. acknowledges support through NSTC grant 111-2112-M-001-064- and 112-2112-M-001-066-.
This work was also supported by the NKFIH NKKP grant ADVANCED 149943 and the NKFIH excellence grant TKP2021-NKTA-64. 
This paper makes use of the following ALMA data: ADS/JAO.ALMA\#2013.1.00163.S, ADS/JAO.ALMA\#2013.1.00387.S and ADS/JAO.ALMA\#2023.1.00907.S. ALMA is a partnership of ESO (representing its member states), NSF (USA) and NINS (Japan), together with NRC (Canada), NSTC and ASIAA (Taiwan), and KASI (Republic of Korea), in cooperation with the Republic of Chile. The Joint ALMA Observatory is operated by ESO, AUI/NRAO and NAOJ.
\end{acknowledgements}

\vspace{-0.7cm}
\bibliographystyle{aa}
\bibliography{ref-abaur.bib}
\begin{appendix}
\section{Observations and tomographies}

\subsection{Observations and data reduction}
\label{app:obs}
Table \ref{tab:obs} summarizes the observed molecular transitions. 
Visibilities obtained in the new Project 2023.1.00907.S (PI:O.Denis-Alpizar) were calibrated using the Casa calibration scripts provided by the ALMA observatory. Calibrated visibilities were then exported into UVFITS data format for imaging and further analysis, using the \textsc{IMAGER} package \footnote{see https://imager.oasu.u-bordeaux.fr}. Spectral resampling and conversion to the LSR velocity frame were done within Casa prior to export in UVFITS format. 

We also analyzed ALMA archival data from Projects 2013.1.00387.S and 2013.2.00163.S \citep{Guilloteau+2016,Dutrey+etal_2017,Simon+2019}. 
We were thus able to derive the proper motions of the source by comparing the apparent positions as a function of time (2012-2024). All observations were then registered to Epoch 2016.0 using the derived proper motion, the nominal position of the source being then R.A. 16:28:13.6979 and Declination -24:31:39.491 \citep[see][]{Guilloteau+2025}. Self-calibration using the continuum data was performed on the most compact configurations only as the source (to first order a bar of apparent size $3 \times 0.3''$) is heavily resolved and does not leave enough flux on most baselines for this purpose.  

\subsection{Imaging}
\label{app:imaging}
After astrometric registration and a small ($\sim 5\%$) flux scale adjustment between epochs \citep[see][for details]{Guilloteau+2025}, all data was then imaged with \textsc{IMAGER} using a common grid, with images of $512 \times  512$ pixels of 0.025$''$ size,  covering a field of view of 12.8$''$. 
Different choices of robust weighting parameters (robust parameter \texttt{MAP\_ROBUST\,=\,2} for high frequency data, and 3 for low frequency ones), were made in order to offer an adequate compromise between angular resolution, sensitivity and dirty beam shape (to minimize in particular the plateau of near sidelobes that occurs when combining two ALMA array configurations). 
A twice larger field of view was used for CO and $^{13}$CO because of contamination by background clouds, although the selected field of view had minor impact on the reconstructed image of the disk itself. 
The typical synthesized beam is $0.27 \times 0.19''$ at PA $90 - 110^\circ$ for the high resolution data, and $0.54 \times 0.51 ''$ at PA $75^\circ$ for the low resolution ones.  Noise levels are respectively $\sim 0.5$\,K and $0.15$\,K at 0.2 km\,s$^{-1}$ spectral resolution.

\begin{table}
\caption{ALMA data.}         
\label{tab:scg}      
\centering                          
\begin{tabular}{|l|c|c|c|c|l|c|}       
\hline              
Band & Science Goals and & Angular  \\
 &  Molecular tracers & resolution  \\
\hline
 B7-new & CS 7-6, HCN 4-3 &  0.18$''$ \\
    & CN 3-2, H$_2$CO & 0.18$''$ \\
        & continuum & 0.18$''$ \\
\hline   
 B6-new & $^{12}$CO, $^{13}$CO, C$^{18}$O 2-1 & 0.2$''$  \\
  & DCN 3-2, $^{13}$CN 2-1, $^{13}$CS 5-4 & 0.2$''$ \\
    & H$_2$CO, c-C$_3$H$_2$, N$_2$D$^+$ & 0.2$''$ \\
      & continuum & 0.2$''$ \\
    \hline
B6-old  & $^{12}$CO, CN 2-1, CS 5-4, H$_2$CO & 0.5$''$ \\
  & continuum & 0.5$''$ \\
\hline                            
\end{tabular} \\
\label{tab:obs}
\end{table}

\begin{figure}
\includegraphics[width=9.0cm]{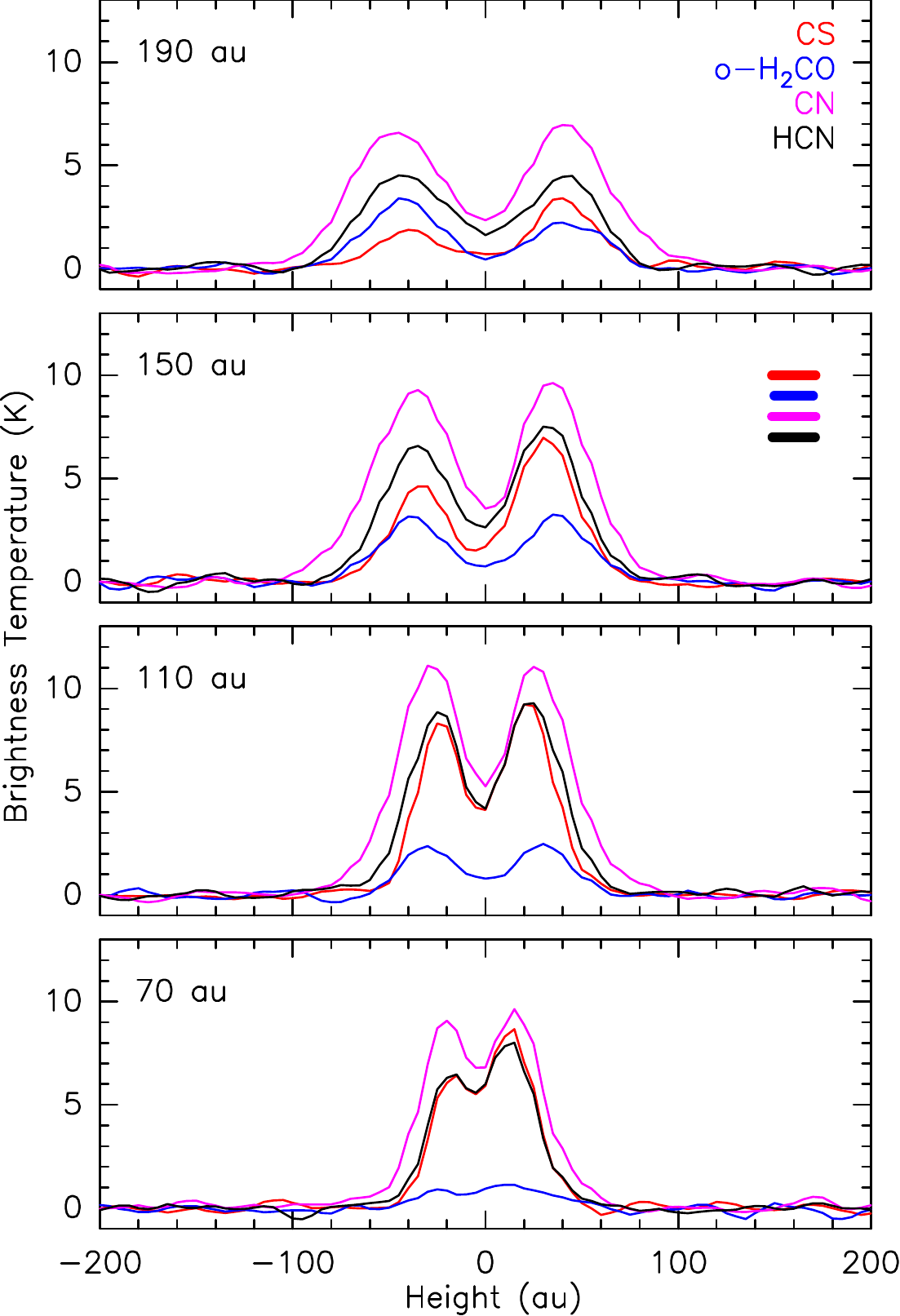}
\caption{Cut in the tomographies at various radii for the brighter lines detected using Band 7. Color bars indicate their respective linear resolutions.} 
\label{Fig:tomo-B7}
\end{figure}

\subsection{Tomographies}
\label{app:description-tomo}

In a rotating disk, the velocity along the line-of-sight projects as the impact parameter therefore each radius appears as a straight line in the Position-Velocity diagram. This allows a direct reconstruction of the 2-D brightness distribution \citep{Dutrey+etal_2017}. 
The tomography has a constant linear resolution as a function of height above the disk plane, but a radius dependent linear resolution decreasing with increasing radius. 
It happens because the Keplerian shear is decreasing with radius compared to the finite local linewidth \citep{Dutrey+etal_2017}. 
Thus, we used the best available spectral resolution to compute the tomography. 
As the original beam is essentially elongated parallel to the mid-plane, the vertical resolution is given by the beam minor axis, and is about 22 au at the assumed distance (to compare with 60 au for the older low resolution data). 

Figure \ref{fig:mean-tomo} shows the tomographies with estimates of the tomographic radial resolution. Note that we did not attempt to correct for line opacity, neither for deviations from the Rayleigh Jeans law. To ease the comparison of the tomographies, Fig.\ref{fig:tomo-super} presents the superimposition of the main molecule emission (in contours) to the $^{13}$CO tomography (in color). 

Fig.\ref{Fig:tomo-B7} presents vertical cuts, revealing 
double Gaussian shapes for vertical distributions. We model these at each radius by fitting two Gaussian, symmetric about the mid-plane in height and width, but of separate intensities:
$$ I(r,z) = I_s(r) \exp\left(-\left(\frac{z+A(r}{W(r)}\right)^2\right) 
+ I_n(r) \exp\left(-\left(\frac{z-A(r}{W(r)}\right)^2\right) . $$
The results of these fits are displayed in Fig.\ref{fig:height-tomo} and Figs.\ref{fig:bright-tomo}-\ref{fig:width} as a function of radius. 
Figure \ref{fig:height-tomo} shows the altitude $A(r)$ and the deconvolved thickness, Fig.\ref{fig:bright-tomo} the peak brightness $\max(I_s(r),I_n(r))$, while Fig.\ref{fig:tomo-ratio} displays the intensity ratio of both sides, $I_s(r)/I_n(r)$. 
The apparent (FWHM) width, $2\sqrt{\log(2)} W(r)$, is displayed in Fig.\ref{fig:width}. 
The intrinsic thickness displayed in Fig.\ref{fig:height-tomo} is recovered from this FWHM
by correcting for the beam size (assuming Gaussian shapes), using the Clean beam minor axis since synthesized beams are elongated almost parallel to the disk plane.

\begin{figure*}
\includegraphics[width=18.0cm,height=7.25cm]{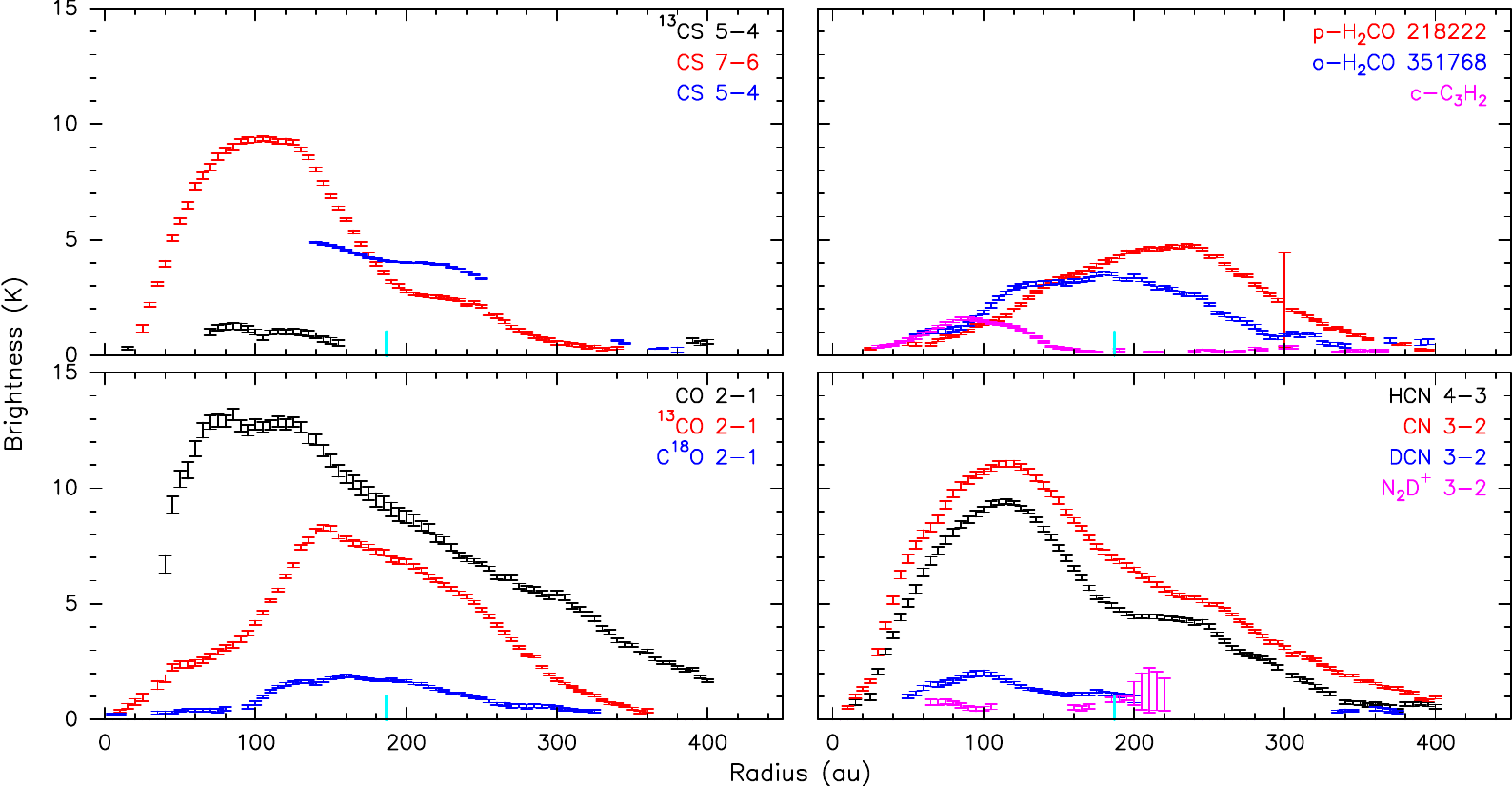}
\caption{Apparent peak brightness $\max(I_s(r),I_n(r))$ as a function of radius.
The cyan bar indicates the dust disk edge.}
\label{fig:bright-tomo}%
\end{figure*}
\begin{figure*}
\includegraphics[width=18.0cm,height=7.25cm]{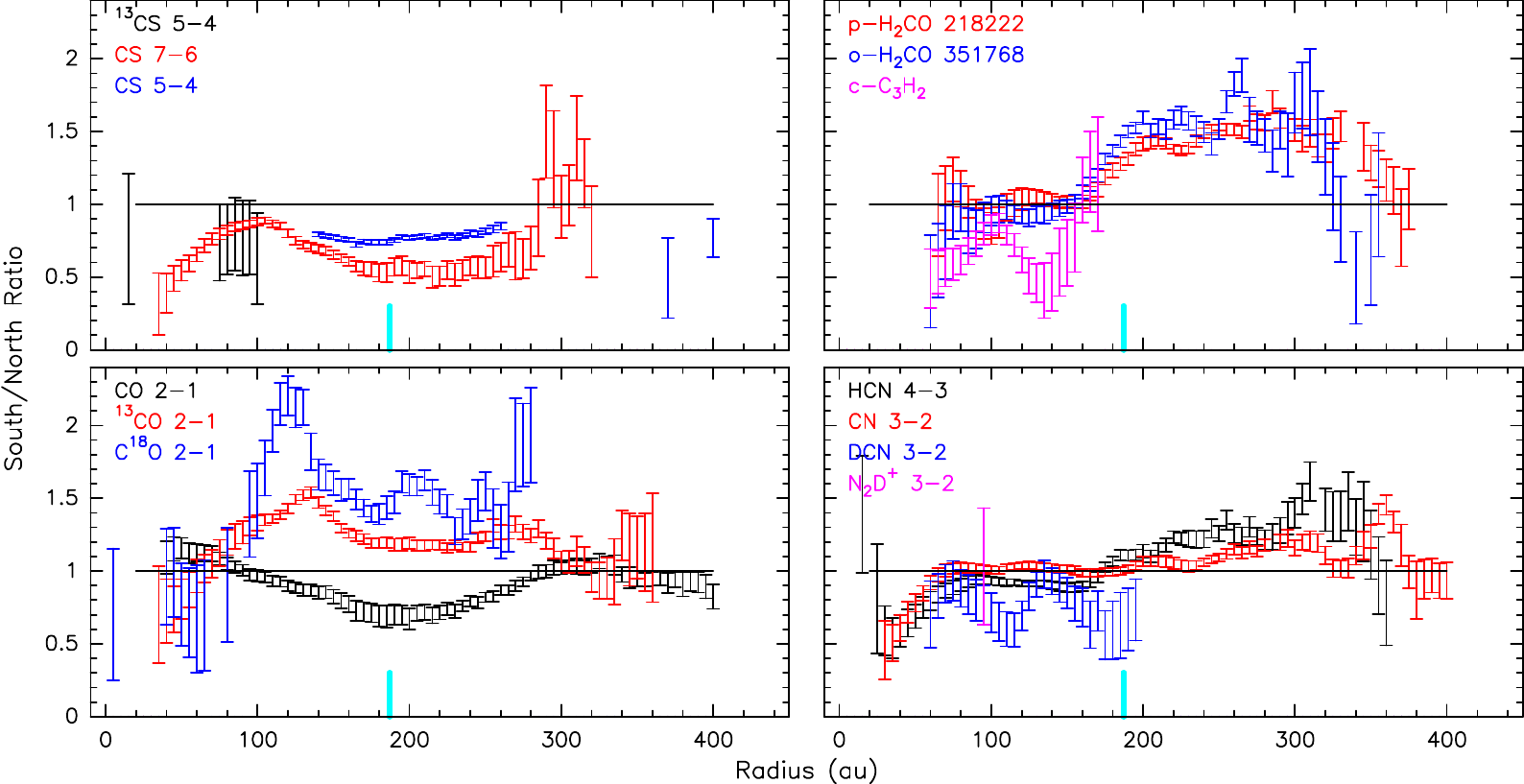}
\caption{South to North layer brightness ratio $I_s(r)/I_n(r)$.
The cyan bar indicates the dust disk edge.}
\label{fig:tomo-ratio}%
\end{figure*}
\begin{figure*}
\includegraphics[width=18cm,height=7.25cm]{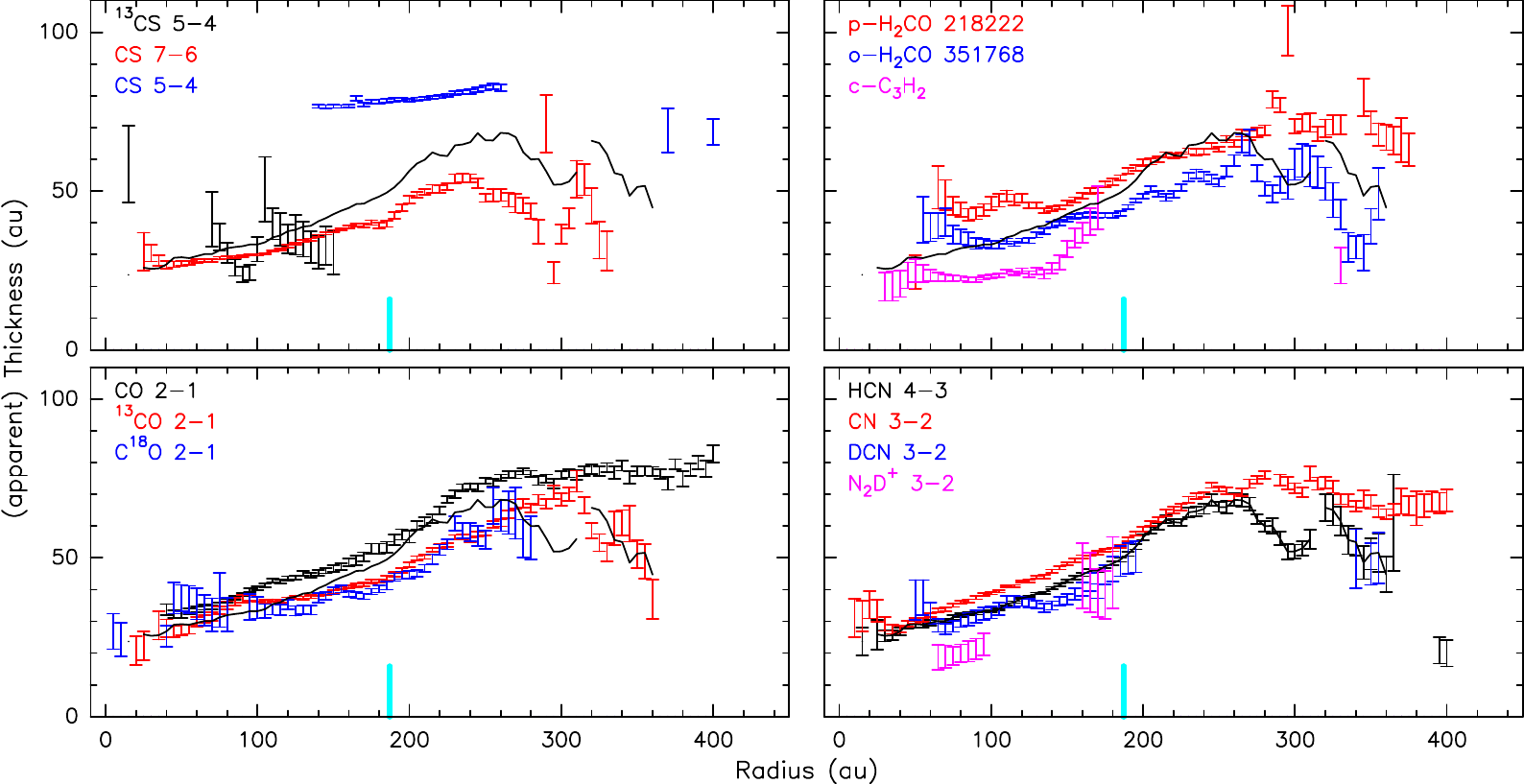}
\caption{Top: Apparent thickness (full width at half maximum) of the molecular layer $2\sqrt{\log(2)} W(r)$, as a function of radius.
The black curve is the HCN(4-3) thickness for comparison.}
\label{fig:width}%
\end{figure*}

\section{Individual molecule fitting}
\label{app:fitting}

\begin{table}[!th] 
\caption{Geometry and velocity for the Flying Saucer disk.}
\label{tab:disk}      
\centering    
\begin{tabular}{|c c|l|}        
\hline              
\multicolumn{2}{|c|}{Geometry} & \multicolumn{1}{c|}{Velocity} \\ 
\hline 
Distance & 120 pc & Stellar Mass  $0.60 \Msun$ \\
$V_\mathrm{LSR}$ & 3.72\,\kms  & $V_\mathrm{100au} = 2.32 \pm 0.01 $\,\kms  \\ 
Inclination & $87.0\pm 0.1^\circ$ &  $e_v$ = 0.5 (Keplerian)\\ 
Orientation & $-3.1\pm 0.1^\circ$ &  \\ 
\hline 
\end{tabular}
\vspace{-1.0ex}
\label{tab:geometry}
\end{table}

For CS, ortho and para H$_2$CO, and C$_3$H$_2$ several transitions were observed, and fitted together, allowing a derivation of the rotation temperature for these molecules. 
Cyclic C$_3$H$_2$ has transitions of its ortho and para forms at identical frequencies: we thus had to assumed an ortho-to-para ratio of 3, the high temperatures limit due to spin statistic, which is
justified since the ground state of ortho-c-C$_3$H$_2$ lies only
2.3 K above that of para-c-C$_3$H$_2$, see 
\citet{Park+2006}.

The two rotational transitions of CN were fitted separately, and an accurate separation of opacity from excitation temperature is possible for each of them because of their complex hyperfine structures, with hyperfine components spanning 1 to 2 orders of magnitude in opacity.
The excellent agreement between both lines indicate that the emission is thermalized.

Only one transition is available for HCN, which has faint hyperfine structure, so that the derived temperature is based on the optically thick region within $r < 150$ au. DCN and N$_2$D$^+$ lines are too optically thin, the temperature had to be assumed. 
We have chosen to present in Table \ref{tab:models} the models assuming the mid-plane temperature (10\,K) found in Paper I, such a temperature being more representative of location of DCN and N$_2$D$^+$ near the mid-plane. 
Note that assuming 17\,K (the derived molecular layer temperature) provides derived surface densities which are only 30\% smaller.

The stronger lines ($^{13}$CO, CS, HCN) are optically thick up to about 150-200\,au, allowing the temperature to be determined. The derived surface density profiles for these molecules is determined from the optically thin region outwards under the assumption that the temperature law remains valid \citep[see][for details, in particular their Fig.4]{Pietu+2007}.

For C$^{18}$O, the temperature was assumed identical to that of $^{13}$CO. 

Both $^{13}$CO and C$^{18}$O lines are affected by background cloud emission, in ranges 3.1 to 4.6,  and 4.1 
to 4.6 km.s$^{-1}$, respectively and the contaminated velocities have been ignored in the fit. For 
$^{12}$CO, we used results from the more elaborate model of \citet{Guilloteau+2025} that uses a vertical temperature gradient and accounts for the background emission.

Our disk model is relatively inappropriate for H$_2$CO, whose emission is dominated
by the outer regions and almost reaches the disk mid-plane. The best fit solutions thus leave some residuals for one of the three bright lines. The ortho-H$_2$CO results seem more influenced by emission from the common molecular layer, in particular the transition at 351.768 GHz, while para-H$_2$CO has a larger contribution from the outer parts. 
\end{appendix}
\end{document}